\documentclass[fleqn,10pt]{wlpeerj}

\usepackage{graphicx}
\usepackage[hidelinks]{hyperref}
\usepackage{xr}
\newcommand{\Anchiornis}{\emph{Anchiornis}}
\newcommand{\Archaeopteryx}{\emph{Archaeopteryx}}
\newcommand{\Confuciusornis}{\emph{Confuciusornis}}
\newcommand{\Jeholornis}{\emph{Jeholornis}}
\newcommand{\Microraptor}{\emph{Microraptor}}
\newcommand{\Microraptorgui}{\emph{Microraptor gui}}

\newcommand{\Sapeornis}{\emph{Sapeornis}}
\newcommand{\Zhongjianornis}{\emph{Zhongjianornis}}
\newcommand{\Zhongornis}{\emph{Zhongornis}}

\newcommand{\Larus}{\emph{Larus}}

\newcommand{\Rhamphorhynchus}{\emph{Rhamphorhynchus}}

\newcommand{\Draco}{\emph{Draco}}
\newcommand{\Supporting}{Supplemental Material}

\newcommand{\mm}{mm}
\newcommand{\mps}{m s$^{-1}$}
\newcommand{\inch}{inch}
\newcommand{\Hz}{Hz}
\newcommand{\cm}{cm}

\usepackage{lineno} % for line numbers during review only

\title{Shifts in stability and control effectiveness during evolution of Paraves support aerial maneuvering hypotheses for flight origin}

\author[1,2]{Dennis Evangelista\thanks{author for correspondence: devangel77b@gmail.com}}
\author[1]{Sharlene Cam}
\author[1]{Tony Huynh}
\author[1]{Austin Kwong}
\author[1]{Homayun Mehrabani}
\author[1]{Kyle Tse}
\author[1,3]{Robert Dudley}
\affil[1]{Department of Integrative Biology, University of California, Berkeley, CA 94720, USA}
\affil[2]{current address: University of North Carolina at Chapel Hill, NC 27599, USA}
\affil[3]{Smithsonian Tropical Research Institute, Balboa, Panama}
\date{\today}


\begin{abstract}
The capacity for aerial maneuvering was likely a major influence on the evolution of flying animals.  Here we evaluate consequences of paravian morphology for aerial performance \citep{Dudley:2011, Smith:1952} by quantifying static stability and control effectiveness of physical models \citep{Evangelista:2014} for numerous taxa sampled from within the lineage leading to birds (Paraves, \citealp{Xu:2011, Gauthier:1985}).  Results of aerodynamic testing are mapped phylogenetically \citep{Maddison:2010, Zhou:2010, Li:2010, OConnor:2011, Cracraft:2004} to examine how maneuvering characteristics correlate with tail shortening, fore- and hindwing elaboration, and other morphological features. In the evolution of Paraves we observe shifts from static stability to inherently unstable aerial planforms; control effectiveness also migrated from tails to the forewings. These shifts suggest that some degree of aerodynamic control and and capacity for maneuvering preceded the evolution of strong power stroke.  The timing of shifts also suggests features normally considered in light of development of a power stroke may play important roles in control.
\end{abstract}

\begin{document}

\flushbottom
\maketitle
\thispagestyle{empty}

%% for line numbers
\setpagewiselinenumbers
\modulolinenumbers[5]
\linenumbers

\section*{Introduction}
Regardless of how aerial behavior originates, once airborne an organism must control \citep{Smith:1952} its orientation and position in order to safely navigate the vertical environment (e.g., directed aerial descent, \citealp{Dudley:2011}). Such abilities are present even in taxa with no obvious morphological adaptation for flight \citep{Munk:2011, Zeng:2013, Cardona:2010, Evangelista:2012}; at low speeds, such as at the start of a fall or jump, inertial mechanisms \citep{Jusufi:2008, Pittman:2013} allow for rolling, pitching, and yawing; as speeds increase (or as appendages grow in area), aerodynamic mechanisms of control can be employed.  Body and appendage configuration and position affect both stability, the tendency to resist perturbations, as well as production of torques and forces for maneuvering (control effectiveness). In the four-winged Early Cretaceous \Microraptorgui, changes in planform, such as alternative reconstruction postures or removal of leg and tail feathers, alter stability and the control effectiveness of appendages \citep{Evangelista:2014}. Furthermore, appendage function can shift entirely according to the aerial environment (e.g. asymmetric wing pronation producing yaw at high glide angle versus roll at low glide angle) or even completely reverse function \citep{Evangelista:2014}. Such results are exciting but are based on a single specimen \citep{Xu:2003}. Stronger conclusions can be drawn from comparative study of several forms within a phylogenetic context. 

One obvious trend in avian evolution is the transition from long tails and feathered legs in early forms \citep{Xu:2003, Hu:2009, Longrich:2006, Christiansen:2004, Zheng:2013, OConnor:2013, Pittman:2013} to later (including extant) forms for which the skeletal tail has fused into a short pygostyle and both asymmetric and symmetric flight feathers are absent from the legs. Functional consequences of this shift for aerial maneuvering remain speculative \citep{Smith:1953, Beebe:1915, Thomas:1997}.   Similarly, changes in the pectoral girdle have been assumed to enhance a powered downstroke \citep{Gauthier:1985, Benton:2005}, but may also have influenced maneuvering by shifting the center of body mass \citep{Allen:2013} or in enabling the production of wing asymmetries. With the exception of \citep{Huynh:2011}, previous aerodynamic studies tend to focus on lift and drag coefficients and glide angles and specific postures \citep{Chatterjee:2007, Alexander:2010, Koehl:2011, Dyke:2013}, with maneuvering only considered rarely and in single taxa \citep{Longrich:2006, Hall:2012, Evangelista:2014}.  
	
To examine these patterns and to test adaptive hypotheses \citep{Padian:2001}, we can use model tests to quantify the effect of shape on static stability and control effectiveness \citep{Evangelista:2014, Koehl:2011, McCay:2001}, using specimens sampled from early paravian \citep{Xu:2011} and avialan \citep{Gauthier:1985} evolution \citep{Zhou:2010, Li:2010, OConnor:2011, Cracraft:2004}.  We focus specifically on static stability and control effectiveness; while lift and drag are expected to be important in flight evolution, they have been addressed adequately in previous literature \citep{Evangelista:2014, Dyke:2013, Koehl:2011, Alexander:2010}.  While the capacity to generate aerodynamic forces for weight support \citep{Dial:2003,Dial:2008,Heers:2014} was almost certainly developing, we consider here the ability to control aerial behavior \citep{Dudley:2011}.  The presence or absence of stability in the various axes and the control effectiveness of the appendages should correlate with changes in major morphological features (shortening of the tail, enlargement of the forewings) to enhance aerodynamic performance, however, previous work has not yet identified what are the patterns.  We hypothesize that stability and control are present early in the evolution of flight \citep{Dudley:2011}; this can be tested by examining the patterns of stability and control.  Morphologically, for an organism able to perform aerial control with only a weak power stroke, we would expect some surface, though not necessarily large, and perhaps with limited force generation due to reduced speed, or feather porosity and flexibility \citep{Nudds:2009, Nudds:2010, Heers:2011}).  We would also predict appendages with sufficient flexibility \citep{Kambic:2014} and inertia \citep{Jusufi:2008}, in combination with aerodynamic forces, to effect maneuvers \citep{Dudley:2011}.  These are all present at the base of the tree of figure~\ref{fig:peerj4}. 

Alternatively, both stability and control may have been absent early in the evolution of flight, only appearing after a strong and bilaterally symmetric power stroke evolved \citep{Dial:2003, Dial:2008, Heers:2014}; or very stable platforms with a lack of effective control mechanisms may be observed.  For these alternatives, morphologically we might expect skeletal features with large surfaces and strong muscle attachments including an early carinate sternum, and shoulders with highly restricted ranges of motion oriented to provide a fundamental stroke oriented to gravity with limited degrees of freedom \citep{Dial:2003, Dial:2008}.  We may also expect rigidly fused/fixed surfaces located posteriorly to provide large stability to offset an inability to generate control. Other then a semilunate carpal assumed to enable a flight stroke \citep{Padian:2001}, these are not observed in the taxa tested here \citep{Benton:2005, Gatesy:2005, Zhou:2010, Turner:2012, OConnor:2012, OConnor:2013}.  Our hypothesis could be falsified: it is possible for some appendage motions or positions to ineffective or unstable, as in the case of certain sprawled postures and leg movements in \Microraptor\ (\citealp{Evangelista:2014}, though these were also anatomically infeasible).

\citet{Smith:1952}, in outlining the potential importance of static stability and control effectiveness, called for measurements to test his assertions: ``If the conclusions of this paper are accepted the study of the remains of primitive flying animals, and even experimental studies on full-scale models of them, will acquire a special importance as throwing new light on the functional evolution of nervous systems \citep{Smith:1952}.'' Subsequent studies of stability have been limited to computational studies with regard to aerodynamic function, etc. \citep{Gatesy:1996, Taylor:2002, Thomas:2001}.  Computational estimates are useful, but when checked against models \citep{Evans:2003} or live animals \citep{Clark:2010} there are occasionally unexpected deviations from aerodynamic theory developed for small angles of attack and airplane-like morphologies.  Therefore, we measured static stability and control by measuring the aerodynamic moments exerted on physical models of different specimens in a wind tunnel, including at large angles of attack.

\section*{Materials and Methods}
\subsection*{Model construction} 
\begin{figure*}[t]
\begin{center} 
\centerline{\includegraphics[width=7in]{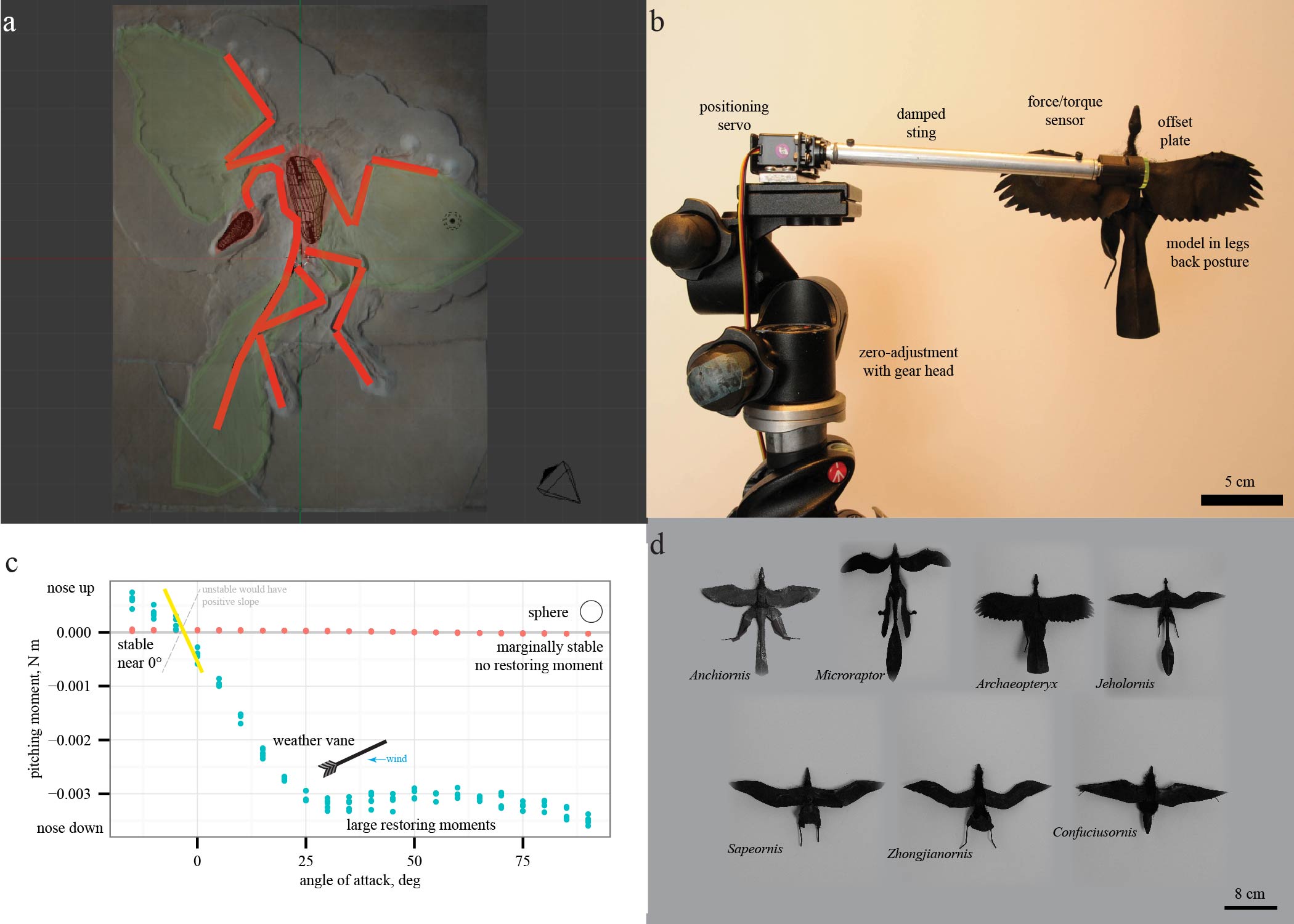}}
%\centerline{\includegraphics{figures/peerj1.pdf}}
\caption{Model construction, testing, and measurement of moments. Models were developed in Blender (a) from fossils (\Archaeopteryx\ shown) and constructed and tested (b) using previous methods \citep{McCay:2001, Koehl:2011, Evangelista:2014}.  For simple cases such as a sphere or a weather vane, the relationship between slope and stability (c) is observed by plotting pitching moments versus angle of attack.  Moments for sphere are not statistically different than zero, indicating marginal stability as expected, further validating the methods. Models for fossil paravians studied are shown in (d). \Anchiornis\ hind limbs rotated out of test position to show plumage for illustration only.}
\label{fig:peerj1}
\end{center}
\end{figure*}

We constructed models (8 \cm\ snout-vent length, figure~\ref{fig:peerj1}) of four extant birds, seven fossil paravians \citep{Xu:2011}, encompassing five avialans \citep{Gauthier:1985}, \Microraptor\ \citep{Xu:2003} and \Anchiornis\ \citep{Hu:2009} (\Supporting, figure~\ref{fig:peerjS1}) using 3D printing. Fossils were selected to sample phylogenies available in 2011 (when the work was done), although an eighth paravian, \Zhongornis\ \citep{Gao:2008}, was later dropped due to questions about its phylogenetic position and because the specimen was identified to be a juvenile. Additional information regarding the morphology of \Microraptor, \Jeholornis, and \Sapeornis\ has become available since the measurements were done; the effect of the new information is addressed in the discussion.  To explore parallel evolution and for calibration, we also constructed models of three pterosaurs, two bats, and two artificial test objects (sphere and weather vane). Construction methods closely followed those of \citep{Koehl:2011, Evangelista:2013, Munk:2011, Zeng:2013}. Solid models were developed in Blender (The Blender Foundation, Amsterdam), closely referencing published photographs of fossils and reconstructions from the literature \citep{Hu:2009, Longrich:2006, Hou:1995, Chiappe:1999, Zhou:2002, Zhou:2003, Xu:2003, Zhou:2003a, Zhou:2010} and casts of \Archaeopteryx\ to match long bone, axial skeleton, and body proportions.  Modeling was also guided by Starling dissections, preserved specimens, and vertebrate anatomy texts \citep{Benton:2005, Liem:2000}.  

Models were printed using a 3D printer (ProJet HD3000, 3D Systems, Rock Hill, SC), then mounted on 26-gauge steel armatures with felt or polymer clay filling in gaps between printed parts where flexibility was needed for repositioning.  Wings were constructed using methods described in \citep{Koehl:2011, Evangelista:2014}; wings were traced from the published reconstructions \citep{Hu:2009, Longrich:2006, Hou:1995, Chiappe:1999, Zhou:2002, Zhou:2003, Xu:2003, Zhou:2003a, Zhou:2010}, printed on paper and cut. Monofilament stiffening was added along feather rachises and the wings were attached to 26-gauge steel limb and tail armatures (scaled from the fossils) using surgical tape (3M, St.~Paul, MN).  This procedure was found in previous work to match more laborious and less repeatable application of manually attached real bird feathers \citep{Koehl:2011} for fixed-wing tests. The~.STL files used to create the models are available for download to researchers wishing to replicate our models.  

\subsection*{Body posture and appendage position} 
\begin{figure}[t]
\begin{center} 
\centerline{\includegraphics{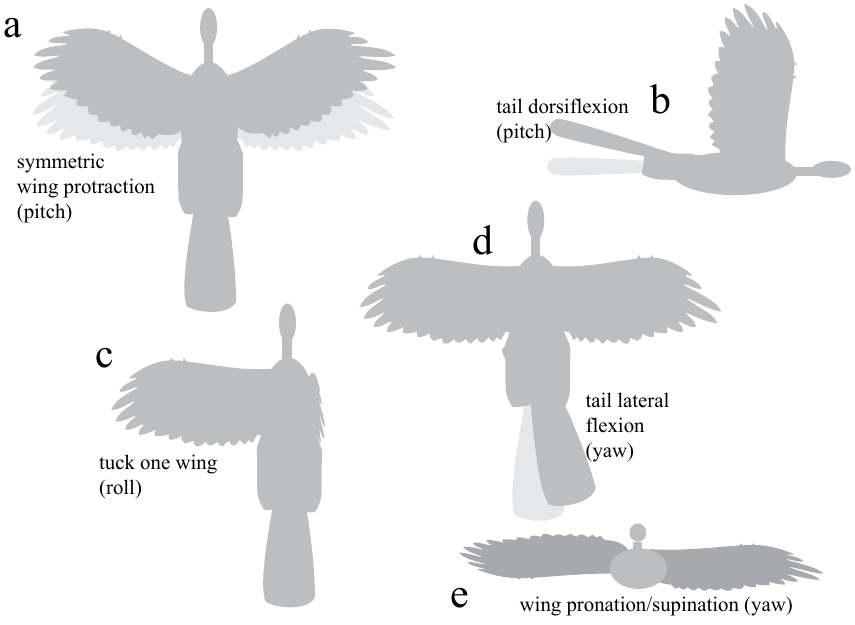}}
\caption{Appendage movements tested to determine control effectiveness.  Light gray indicates baseline posture, dark gray indicates appendage deflection.  Appendage movements were selected based on those observed to be effective in previous work \citep{Evangelista:2014}, including (a) symmetric wing protraction (e.g. wing sweep to $\pm 45^{\circ}$); (b) tail dorsiflexion to $\pm 15^{\circ}$; (c) tucking of one wing; (d) tail lateral flexion to $30^{\circ}$; and (e) asymmetric wing pronation/supination to $15^{\circ}$ (e.g. left wing pitched down, right wing pitched up).}
\label{fig:peerj2}
\end{center}
\end{figure}

Fossil paravian models were reconstructed with wings spread and legs extended back \citep{Xu:2004,Evangelista:2014} (figures~\ref{fig:peerj1} and \ref{fig:peerj2}).  While alternative postures have been considered (specifically in \Microraptor; \citealp{Xu:2003, Chatterjee:2007, Nova, Alexander:2010, Hone:2010, Koehl:2011, Habib:2012, Hall:2012, Evangelista:2014, Dyke:2013}; including some now considered infeasible), the aim of this study was to examine maneuvering within several species rather than the posture of one, and the legs-back posture is seen in extant birds as well as supported for the fossils \citep{Xu:2004, Nova}. For control effectiveness, we tested fixed static appendage movements previously identified as being aerodynamically effective \citep{Evangelista:2014,Evangelista:2013}: asymmetric wing pronation and supination, wing tucking, symmetric wing protraction and retraction, and dorsoventral and lateral movements of the tail (figure~\ref{fig:peerj2}). The angular extent of each movement tested is shown on figure~\ref{fig:peerj2}. To avoid confusion, we present data for specimens as described with all appendages present; artificial manipulations (such as removal of tail and leg surfaces) were discussed in \citep{Evangelista:2014, Evangelista:2013}. 
%We invite proponents of other postures to repeat our methods for their chosen postures.

Models were mounted at the estimated center of mass (COM) for the baseline body posture.  The estimate was formed in Blender assuming a uniform density for the posed model, as in \citep{Allen:2013}.  While we did not duplicate the same sensitivity analyses as \citep{Allen:2013}, we recognize that the COM estimate could vary up to 3-5\% of the body length, or by smaller amounts for the variable load associated with appendage movements; this uncertainty is usually within the bounds of coefficient estimates identified as marginally stable.  Mass and wing loading of the organism, while important for estimating speed from lift and drag coefficients, do not directly affect the nondimensional coefficients presented here.  

\subsection*{Wind tunnel testing} Wind tunnel testing used previous methods \citep{Evangelista:2014}, with a six-axis sensor (Nano17, ATI, Apex, NC) mounted to a 0.5 \inch\ (12.7 \mm) damped sting exiting the model downwind at the center of mass (figure~\ref{fig:peerj1}). In some measurements, a 2 \mm\ steel extension rod or a 3 \mm\ acrylic plate were used to avoid geometric interferences and to keep the sting several diameters away and downstream of aerodynamic surfaces.  The sensor was zeroed for each measurement, eliminating model deadweight effects.  Models were tested in an open-circuit Eiffel-type wind tunnel with an 18$\times$18$\times$36-inch (45.7$\times$45.7$\times$91.4 cm) working section (Engineering Laboratory Design, Lake City, MN).  Testing at 6 \mps\ resulted in a Reynolds number of $\sim$32,000 for all models, matching full scale for \Anchiornis, \Archaeopteryx, \Sapeornis, \Zhongjianornis, and \Confuciusornis.

Under the test conditions, the aerodynamic coefficients of interest are reasonably constant with Reynolds number, $\mbox{Re}=UL/\nu$, where $L$ here is the snout-vent length and $\nu$ is the kinematic viscosity of air \citep{Evangelista:2014,Evangelista:2013}.  Early in the evolution of animal flight, organisms likely flew at moderate speeds and high angles of attack \citep{Evangelista:2014, Dyke:2013} where flows appear like bluff body turbulent flows (in which coefficients are largely independent of $\mbox{Re}$, for $10^3<\mbox{Re}<10^6$).  In previous work \citep{Koehl:2011, Evangelista:2014}, we performed a sweep of wind tunnel speed, to examine $\mbox{Re}$ from 30,000 to 70,000, to validate that scale effects were not present.  As additional support for this approach, tests for maneuvering bodies are nearly always tested at well below full scale $\mbox{Re}$, e.g. the largest US Navy freely-maneuvering model tests are well below $\frac{1}{3}$-scale. Our methods were also previously benchmarked using model tests at full scale $\mbox{Re}$ of gliding frogs \citep{Emerson:1990b, McCay:2001} (repeated for comparison), \Draco\ lizards, Anna's Hummingbirds in glide and extreme dive pullout maneuvers, hummingbird body shapes in hovering \citep{Sapir:2012}, and reduced-scale tests of human skydivers compared to actual data \citep{Cardona:2010, Evangelista:2012}; while at $\mbox{Re} \sim 1000$, our modeling methods have been benchmarked against extant winged seeds. Perching aerial robots, developed to test control algorithms, have shown good agreement between fully 3D robots and flat plate models with the same planform \citep{Roberts:2009, Hoburg:2009, Tangler:2005}.  Results \citep{Evangelista:2014} for lift and drag coefficients using our method agreed with those for full-scale \Microraptor\ models in the other modeling tests \citep{Dyke:2013, Alexander:2010}; our \Jeholornis\ model was at lower Re than \Microraptor\ and is of similar shape.
	
Sensor readings were recorded at 1000 \Hz\ using a data acquisition card (National Instruments, Austin, TX) \citep{Evangelista:2014}. The sting was mounted to a servo (Hitec USA, Poway, CA) interfaced to a data acquisition computer, using an Arduino microcontroller (SparkFun, Boulder, CO) and specially written code in Python and R \citep{R:2013}, to automate positioning and measurement of windspeed and force/torque.  Raw measurements were rotated to a frame aligned with the wind tunnel and flow using the combined roll, pitch, and yaw angles by multiplication with three Euler rotation matrices; translation from the sensor to the model COM was also included.  Transformed measurements were averaged over a one-minute recording.  We then computed non-dimensional force and moment coefficients, static stability coefficients, and control effectiveness \citep{McCay:2001, Evangelista:2014, McCormick:1995}. Three series, varying pitch, roll, and yaw, were conducted at 5$^{\circ}$ increments.  Using the automatic sting, we obtained 13,792 measurements, with at least five replicates for 18 models in 247 total positions: generally 5 each in pitch (88 total), 2 each in roll for two angles of attack (69 total), and 3 each in yaw for two angles of attack (92 total). Test positions are indicated in figure~\ref{fig:peerj2}. 

Static stability was measured by examining the sign of the slope $\partial C_m/\partial \alpha$ (positive slope is unstable, negative stable, zero marginally stable, see figure~\ref{fig:peerj1}c) of the non-dimensional pitching moment coefficient $C_m$ near fixed points as the body was subjected to small deflections $d\alpha$ \citep{Evangelista:2014, McCay:2001, McCormick:1995}:
\begin{equation}
\mbox{pitching moment } M = 0.5 \rho U^2 C_m \lambda S
\end{equation}
where $U$ is tunnel speed, $\lambda$ is the snout-vent length, and $S$ is planform area.  Control effectiveness ($\partial C_m/\partial\delta$, \citep{Etkin:1996, McCay:2001, Evangelista:2014}) was measured by deflecting appendages (figure~\ref{fig:peerj2}) by an amount $d\delta$ and examining the change in pitching moment coefficient.  Both are unitless (rad$^{-1}$).  Roll and yaw were similarly calculated from series of varying roll angles or headings, respectively, with the static stability and control effectiveness partial derivatives taken for the roll ($0.5\rho U^2 C_r \lambda S$) and yaw ($0.5 \rho U^2 C_y \lambda S$) moment coefficients. 
 
A first-order estimate of maneuvering is obtained from considering the two together and a biomechanical trade-off is apparent: a stable object can resist perturbations from the environment with minimal control effort but will also have difficulty in changing direction (which might be necessary to accomplish aerial righting, navigate in cluttered forests, seek resources or avoid predators) \citep{Smith:1952,Dudley:2011,Taylor:2002}.  The metrics underestimate maneuvering in very dynamic cases (high advance ratio flapping or where second-order damping terms become important; \citealp{Sachs:2005}), but are adequate for quasi-static maneuvers. Locomotion is a complex task, and passive stability is often exploited where possible to reduce control effort; conversely, passive instability may be exploited in extreme (and likely elective) maneuvers. The absence of stability, coupled with the presence of large control effectiveness, could be used to infer the presence of strong closed-loop neuromuscular control.  The absence of control effectiveness suggests a lack of control, as even with feedback an ineffective surface cannot generate the necessary forces and torques.  Thus, while the full control abilities of an extinct form are difficult if not impossible to fully enumerate, the simple metrics here provide a useful proxy.
	
\subsection*{Phylogenetic comparisons} A Nexus file without branch lengths, available in the \Supporting, was assembled from published phylogenies \citep{Zhou:2010, Li:2010, OConnor:2011, Cracraft:2004} of the study taxa.   While revisions to the phylogenetic relationships have been discussed \citep{Xu:2011, Godefroit:2013, Turner:2012, OConnor:2013}, they do not appear to alter the patterns in stability and control effectiveness; trees from \citep{Xu:2011, Godefroit:2013, Turner:2012, OConnor:2013} are included in the \Supporting.  Mapping of discrete maneuvering traits as outlined in \cite{Padian:2001} was performed in Mesquite \citep{Maddison:2010} with the built-in ancestral state reconstruction routines using unordered parsimony. Aerodynamic measurements were coded into a matrix giving eight discretized stability values (stable, marginal, unstable, coded according to if the 95\% confidence interval of measurements includes zero or not) and 12 discretized control effectiveness values.  The discretized control effectiveness values were obtained from the measurements by thresholding based on the moment necessary to overcome measured weather vane stability ($dC/d\delta>0.09$ was coded as effective; $<0.09$ coded as ineffective), or equivalently, to cause a displacement of the center of aerodynamic pressure of about 10\% of total length.

\section*{Results}
\begin{figure}[ht]
\begin{center}
\centerline{\includegraphics{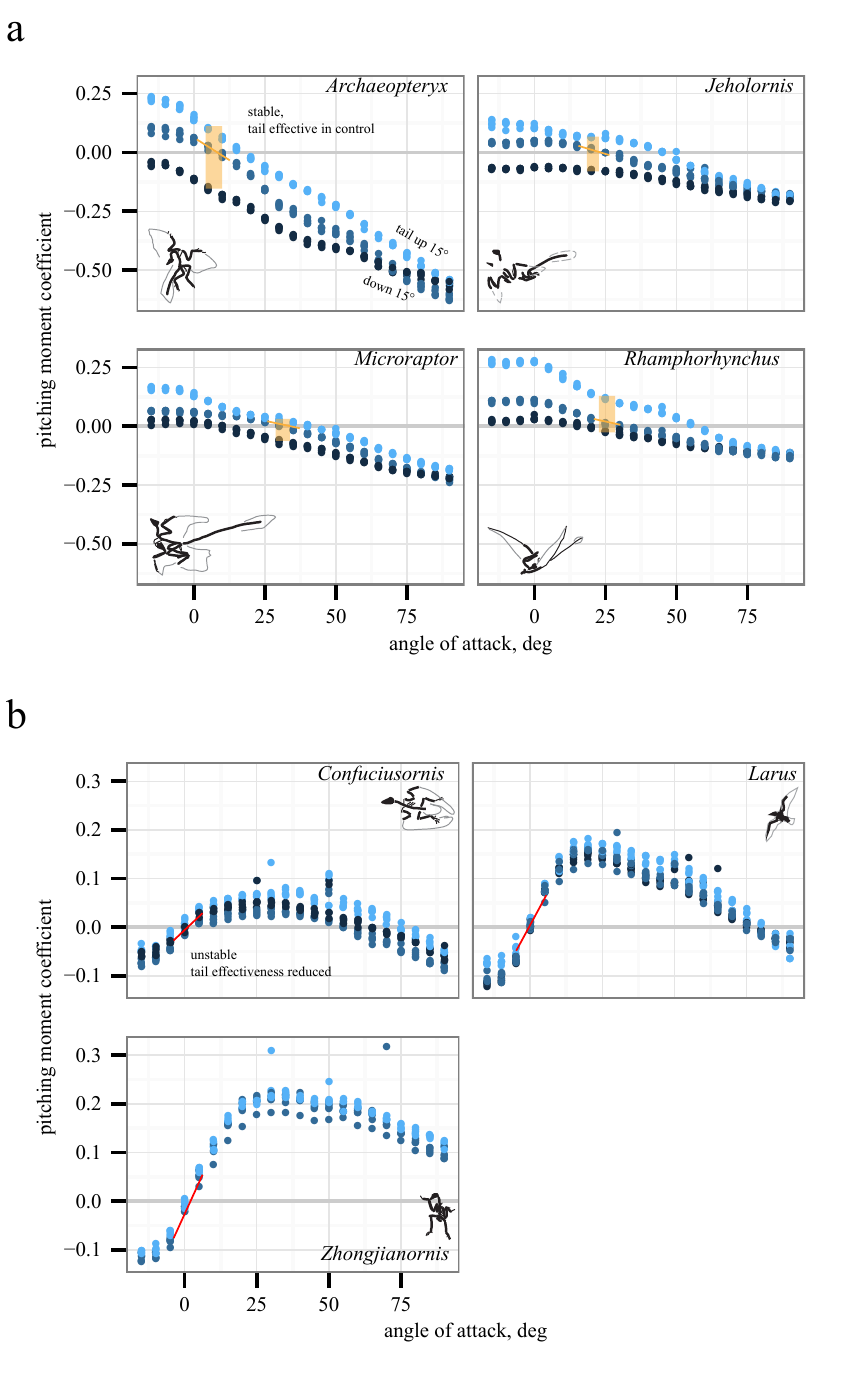}}
\caption{Representative aerodynamic measurements for pitching stability and control effectiveness.  Long-tailed taxa (a) have a stable equilibrium point at 10-25$^{\circ}$ (yellow line) and the tail is effective in generating pitching moments at low angles of attack (pale yellow box indicates measurable moments for given tail deflections).  In short-tailed taxa (b), including extant \Larus, the equilibrium point at 0-5$^{\circ}$ is unstable (red line) and the tail control effectiveness is reduced (no measurable moments for the given tail deflections).  One example (\Rhamphorhynchus) drawn from pterosaurs illustrates similar possibilities in a phylogenetically distant taxon.}
\label{fig:peerj3}
\end{center}
\end{figure}

\begin{figure}[ht]
\begin{center}
\centerline{\includegraphics[width=3.42in]{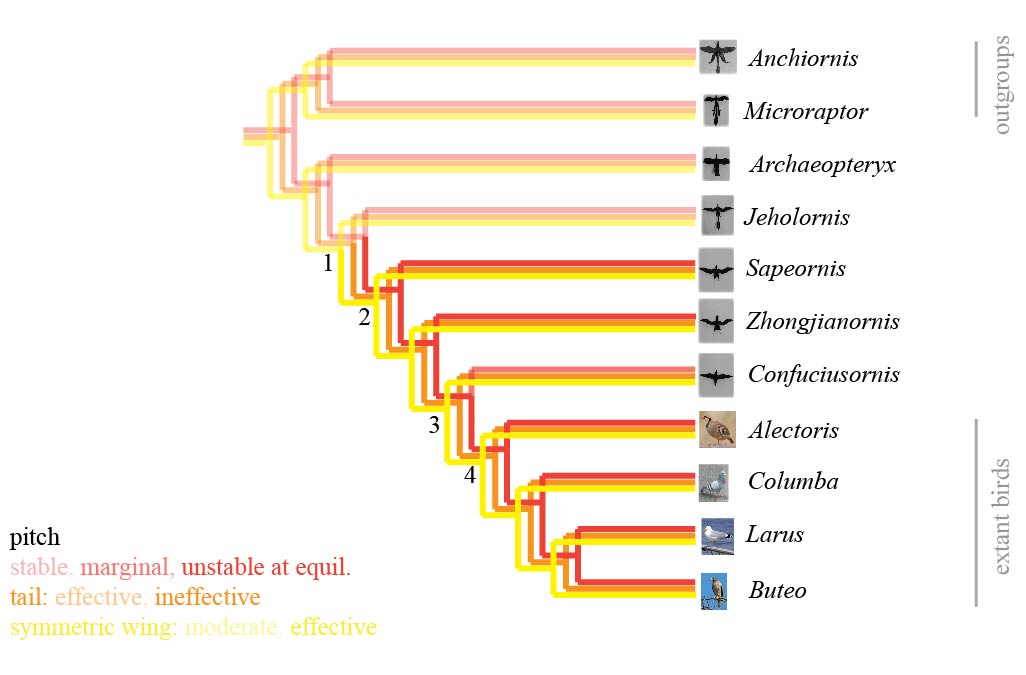}}
%\centerline{\includegraphics{figures/peerj4-cartoons.pdf}}
\caption{Evolution of pitch stability and control effectiveness.  Pitching stability is plotted in red hues, indicating stable (pale), marginally stable (medium), and unstable (solid).  Control effectiveness of the tail in generating pitching moments is plotted in orange hues, indicating large control effectiveness (pale) or reduced control effectiveness (solid). Control effectiveness of symmetric wing protraction/retraction is plotted in yellow hues indicating large (pale) or reduced (solid).  Consilience among the three traits (nodes 1-2) indicates that early in the evolution of Paraves, taxa are stable with a large degree of pitch control from the tail; later taxa are unstable, and control has migrated from the now reduced tail, to the wings, which become larger and develop skeletal features that would enhance control and the production of left-right and fore-aft asymmetries. Nodes 1-4 are discussed further in the text.}
\label{fig:peerj4}
\end{center}
\end{figure}

\begin{figure}[ht] 
\begin{center}
\centerline{\includegraphics[width=3.42in]{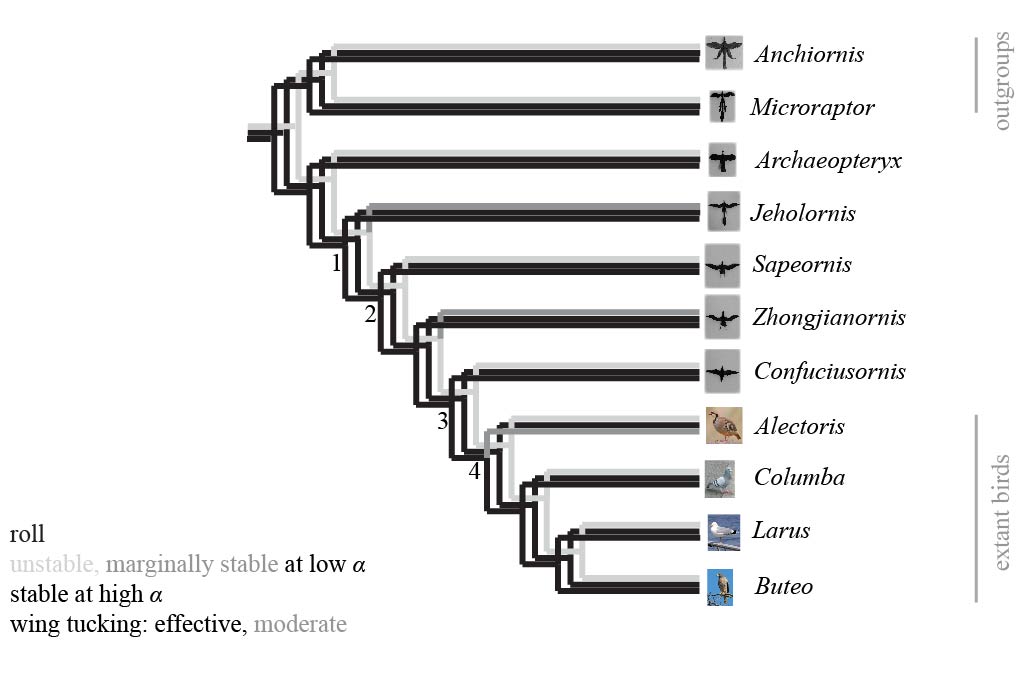}}
%\centerline{\includegraphics{figures/peerj5-cartoons.pdf}}
\caption{Evolution of roll stability and control effectiveness.  Characters shown are stability at low (15$^{\circ}$) angle of attack (mostly unstable due to symmetry; \Sapeornis\ marginal); stability at high (75$^{\circ}$) angles of attack (all stable); and control effectiveness of asymmetric wing tucking in roll (always effective).  As animals developed the ability to fly at reduced body angles of attack, more active control of roll would have been necessary, at first perhaps using inertial modes of the tail, but subsequently augmented with the forewings.}
\label{fig:peerj5}
\end{center}
\end{figure}

\begin{figure}[ht]
\begin{center}
\centerline{\includegraphics[width=3.42in]{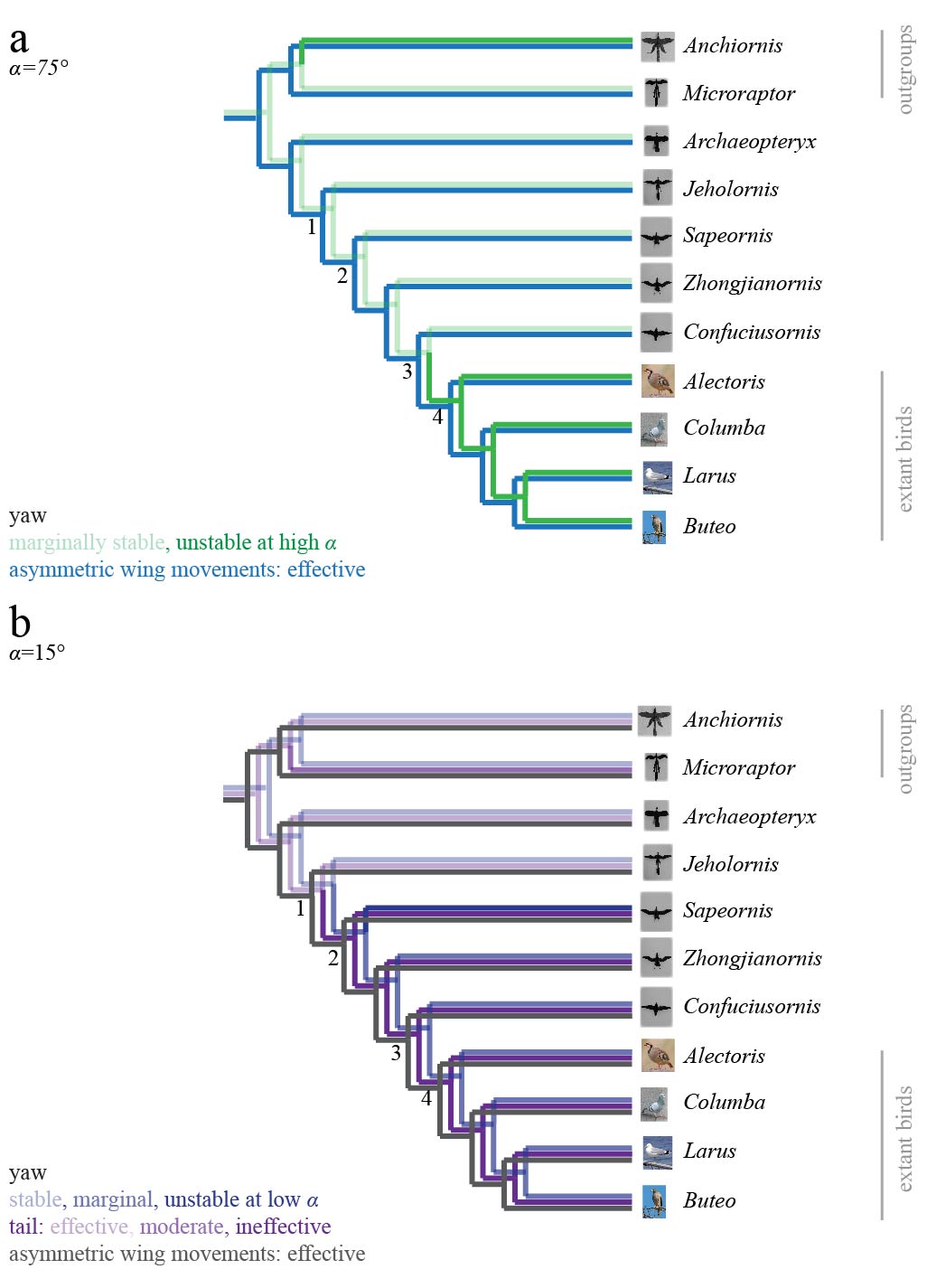}}
%\centerline{\includegraphics{figures/peerj6-cartoons.pdf}}
\caption{Evolution of yaw stability and control effectiveness. At high (75$^{\circ}$) angles of attack (a), taxa are mostly marginally stable as might be expected for high angles (e.g. at 90$^{\circ}$ angle of attack all forms are marginal).  Asymmetric pronation and supination of the wings are always effective in generating yaw at high angles of attack.  At low (15$^{\circ}$) angles of attack (b), by contrast, long-tailed taxa are stable and can control yaw with the tail.  As tails reduce in size (nodes 1-2), taxa become unstable in yaw at low angles of attack and lose the ability to control yaw with the tail.  However, asymmetric movements of the wings are effective in producing yaw throughout the evolution of this clade, and control would thus have shifted from the tail to the forewings, paralleling the shifts seen in pitch.}
\label{fig:peerj6}
\end{center}
\end{figure}

Representative aerodynamic measurements for pitching stability and control effectiveness are given in figure~\ref{fig:peerj3} for six paravians and one pterosaur.  Tables of all aerodynamic measurements are provided in the \Supporting.  All aerodynamic measurements were coded into both discretized and continuous character matrices (see \Supporting), which were then mapped onto a phylogeny (assembled from \citealp{Zhou:2010, Li:2010, OConnor:2011, Cracraft:2004}) to examine the evolution of static stability and control effectiveness. The discretized character states for pitch, roll, and yaw are shown in figures~\ref{fig:peerj4}-\ref{fig:peerj6}.  For these figures, color indicates character while the alpha transparency (how faded or solid the line appears) indicates the character state.  The trees are shown to explicitly illustrate the patterns of aerial maneuvering for several different character states within a phylogeny \citep{Padian:2001}.  Alternate phylogenies \citep{Xu:2011,Godefroit:2013,Turner:2012,OConnor:2013} are given in the \Supporting, but do not alter the patterns seen. Each trait is subject to uncertainties of measurement (see \Supporting), the limitations of ancestral state reconstructions with unordered parsimony, and uncertainty in the phylogenies, however, in the aggregate the results show consilience (later taxa lines are solid) among pitch and yaw traits as discussed further below.

\section*{Discussion}

\subsection*{Additional information and new reconstructions}
The aerodynamic tests described here were performed in the fall of 2011; since then, more information has become available leading to more detailed reconstructions and phylogenies being proposed \citep{Li:2012, OConnor:2012, OConnor:2013, Zheng:2013, Foth:2014}.  While we are not able to perform new measurements, we can theorize about the effect of these newer reconstructions.  \citet{Foth:2014} confirmed \citeauthor{Longrich:2006}'s \citeyearpar{Longrich:2006} finding of leg feathers and does not alter our results.  \citet{Li:2012} provide a detailed tail reconstruction of \Microraptor\ tail plumage including a graduated tail shape with long, midline feathers; we estimate that this tail morphology may have slightly different values for stability and control effectiveness but the overall presence/absence pattern we observed here would be unchanged.  \citet{OConnor:2012, OConnor:2013} provide further information on plumage for \Jeholornis; while our forewing reconstruction appears adequate, we lack the proximal fan of the ``two-tailed'' morphology identified in newer specimens.  The additional contribution of the proximal fan is hard to estimate, and should be considered in future work; however, its forward position suggests a small moment arm and therefore a small effect. This is further supported by the marginal stability and small control effectiveness of unrealistically sprawled legs, which were also near the center of mass, observed in \Microraptor\ models \citep{Evangelista:2014}. \citet{Zheng:2013} identify distal leg feathers and a long/broad tail in new specimens of \Sapeornis, while \citet{Turner:2012} revise the position of \Sapeornis\ to be more basal than \Jeholornis.  As tested, a more basal position for \Sapeornis\ complicates the interpretation of our findings, thus we include a mapping onto such a tree in the \Supporting.  However, taken together, the additional leg and tail plumage described in \citep{Zheng:2013} and the more basal position proposed in \citep{Turner:2012, OConnor:2013} would maintain the patterns we see here, shifting one node earlier in the case of pitch (from node 2 to node 1).     

\subsection*{Patterns in longitudinal stability and control}
Long-tailed taxa (figure~\ref{fig:peerj3}a) show a stable equilibrium point and the tail is effective in generating pitching moments, whereas short-tailed taxa (figure~\ref{fig:peerj3}b) were unstable and had reduced control effectiveness of the tail.  Notably, the same pattern (i.e., downward sloping $C_m$ versus $\alpha$) is seen consistently in two early representatives of the Avialae, in a long-tailed pterosaur (\Rhamphorhynchus), and in the paravian Early Cretaceous dromaeosaur \Microraptor, suggesting that these patterns more likely derive from shape and aerodynamics than from immediate ancestry. 

The study taxa show progressive tail loss as well as loss of leg-associated control surfaces along with a concomitant increase in forewing size.  Changes in stability and control effectiveness here (as well as manipulations in which appendage surfaces were removed with all else held constant \citep{Evangelista:2014}) reflect these morphological changes. In pitch (figure~\ref{fig:peerj4}), taxa shift from being statically stable ancestrally to subsequently being unstable (and thus requiring active control, or possibly damping from flapping counter-torques; figure~\ref{fig:peerj4} red line).  Control effectiveness concomitantly migrates from the ancestrally large and feathered tail (orange line) and legs to the increasingly capable forewings (yellow line), which become relatively larger, gain larger muscle attachments and gain skeletal features and stiffness proximally \citep{Benton:2005} that would improve production of left-right and fore-aft kinematic asymmetries needed for control. Distally, bone loss, joint fusion and use of ligaments to reduce degrees of freedom \citep{Benton:2005} would have enabled mechanical feedback and tuned mechanisms as flapping developed, enabling neuromuscular control effort to be focused on dealing with increasing overall flight instability and active control.  

Transition to forewing control correlates with a significantly enlarged humeral deltopectoral crest \citep{Zhou:2010} and occurs amid progressive acquisition of a fully ``avian'' shoulder morphology \citep{Turner:2012}.  In addition, the sternum is changing from ossified in \Microraptor\ and \Anchiornis\, through varying degrees of loss or ossification without fusion (around nodes 1-2), to ossification, with later fusion (node 3) and development of a carinate keel (node 4).  Concomitantly, the tail becomes much reduced into a pygostyle (Figure~\ref{fig:peerj4}, node 2) with increased mechanical stiffness \citep{Pittman:2013}, which, combined would have decreased the moments the tail could produce and eliminated inertial mechanisms.  Other synapomorphies appear at node 4 (figure~\ref{fig:peerj4}), including a strut-like coracoid and triosseal canal \citep{Benton:2005}. Whereas the latter features (node 4) feature in power production, the timing of the former features (nodes 1-2) appears consistent with enhanced forewing control effectiveness.  Ontogenetic tests \citep{Evangelista:2013} show 4-day post hatching Chukar Partridge are capable of extreme maneuvers (rolling and pitching 180$^{\circ}$) before strong development of a carinate sternum, suggesting this interpretation is correct.

\subsection*{Roll and yaw control at high angle of attack is present early}
In roll (figure~\ref{fig:peerj5}), taxa were stable at high angles of attack, but either unstable or marginally stable at low angles of attack; large asymmetric wing movements (i.e., wing tucking) were always effective in creating substantial rolling moments early, well before development of a power stroke. Also, as animals developed the ability to fly at lower angles of attack, active control of roll would have become necessary, perhaps via inertial modes of the tail (only available before node 2) or legs, and later augmented with the forewings as they became larger and more capable of left-right asymmetry during full force production (carinate sternum, node 4). In all channels, the presence of control effectiveness early is contrary to assertions that aerodynamic functions in early paravians are unimportant \citep{Foth:2014}. Wing movements with high control effectiveness change during evolution in a manner consistent with predicted shoulder joint mobility (criterion 1 of \citealp{Gatesy:2005}). 

In yaw, most taxa at high angles of attack (figure~\ref{fig:peerj6}a, green lines) were marginally stable as might be expected from symmetry.  Taxa with long tails were also stable at low angle of attack (figure~\ref{fig:peerj6}b, purple line), in agreement with computational predictions for similar shapes \citep{Sachs:2007}, but as tails are reduced, yaw stability becomes marginal and control migrates from the tail (violet line) to the wings (as in pitch control, gray line).  Asymmetric wing pronation and supination (figures~\ref{fig:peerj6}a and \ref{fig:peerj6}b) was effective in generating yawing moments in all taxa, suggesting maneuverability in yaw early in bird evolution.  As the tail becomes shorter, flight becomes marginally stable or unstable (node 3), and control effectiveness must migrate (as in pitch control) from the shortening tail to the enlarging forewings. In yaw as in roll, it is possible that a carinate sternum (node 4) enables more capable left-right asymmetry during full force production in extant birds.  

Stability in roll and in yaw shifts with angle of attack, which may serve as a proxy for glide angle or the steepness of descent.  At high angles of attack, roll is the more stable axis, whereas stability in yaw is greater at low angles of attack.  Forewing asymmetrical movements created substantial yawing moments in all taxa, whereas forewing symmetrical movements were only effective in later taxa.  The presence of such stability shifts is more consistent with a transition from high glide angles to lower glide angles, as predicted by an aerial maneuvering hypothesis \citep{Dudley:2011}; they are inconsistent with a fundamental wing stroke with fixed orientation to gravity.  Increased stiffness of the shortening tail \citep{Pittman:2013} would also enable high force production at occasional, critical moments of high angle of attack flight, such as during landing.  The stability and control effectiveness patterns we observed illustrate a control system that works at steep angles in ancestral taxa, shifting to one optimized for low angles in derived taxa including extant birds.

% The reviewer is incorrect.  Sachs deals with dynamic stability in yaw, not static or passive stability.  The dynamic stability effects Sachs identify stem from rotational damping and are only possible for a shape that is marginally stable in static stability, which ours were.  Our results agree with Sachs in this respect. 
 
\subsection*{Maneuvering and the evolution of flight}
The findings suggest that the capacity for maneuvering characterized the early stages of flight evolution \citep{Dudley:2011}, before forewings with a power stroke fully evolved. Although early paravians may have been limited to tight coupling of vertebral and retricial movement \citep{Gatesy:1996}, overall gross movement of the large tail of early paravians yielded high aerodynamic control effectiveness and the body possessed some degree of stability.  Combined with likely dynamic forces and torques generated by either tail whipping at the root \citep{Jusufi:2008, Pittman:2013} or mild asymmetric or symmetric forewing flapping (flapping limited by less robust skeletal and feather morphology or porosity), this suggests that the ancestral organisms were still capable of controlled aerial behaviors at high angles of attack (figures~\ref{fig:peerj4}-\ref{fig:peerj6}).  Gradual evolution of improved maneuvering ability (increased control effectiveness, reduced stability) via consistent aerodynamic mechanisms is consistent with a continuum of aerial behaviors ranging to full aerial (\citealp{Dudley:2011}; criteria 3 and 4 of \citealp{Gatesy:2005}).  The staggered acquisition of certain morphological characters (e.g. sternum ossification; pygostyle) is consistent with aerial maneuvering evolving in an incremental, rather than quantum, manner.  Subsequent shifts in control would be consistent with more shallow glides facilitated by incipient wing flapping, which may have served initially in control but then ultimately became the power stroke characteristic of modern birds.  Incipient flapping may thus have become elaborated as a control response \citep{Smith:1952} to instabilities (figure~\ref{fig:peerj4} node 1; figure~\ref{fig:peerj6}a node 3; figure~\ref{fig:peerj6}b node 1) demonstrated here.  Body center of mass was migrating forward \citep{Allen:2013}, but this is coupled with loss of large posterior surfaces (long tail and leg flight feathers) and coincidence of the wing center with the COM.  Active control was thus required as static stability was reduced and eventually lost, and associated forewing movements would also have enhanced aerodynamic force production and provided a means for inertial attitude adjustment. Once the transition to wing-mediated maneuverability and control began, larger surfaces and increased musculature would have facilitated dynamic force production for weight offset via the power stroke characteristic of modern birds.

\section*{Acknowledgments}
We thank Y.~Munk, Y.~Zeng, E.~Kim, M.~Wolf, N.~Sapir, V.~Ortega, S.~Werning, K.~Peterson, J.~McGuire and R.~Fearing for their advice and assistance.  We than the Berkeley Undergraduate Research Apprentice Program (URAP) and the help of G.~Cardona, C.~Chun, M.~Cohen, E.~Guenther-Gleason, V.~Howard, S.~Jaini, F.~Linn, C.~Lopez, A.~Lowenstein, D.~Manohara, D.~Marks, N.~Ray, A.~Tisbe, F.~Wong, O.~Yu and R.~Zhu.  The manuscript was improved using comments from nine anonymous reviewers.  We also thank T.~Libby and the Berkeley Centre for Integrative Biomechanics in Education and Research (CIBER) for use of a force sensor and 3D printer. 
%DE was supported by an NSF Minority Graduate Research Fellowship, UC Chancellor's Fellowship, and NSF Integrative Graduate Education and Research Traineeship (IGERT) \#DGE-0903711.  TH was supported by the University of California Museum of Palaeontology (UCMP).

\bibliography{evangelista-peerj}

\end{document}